# Theoretical analysis and kinetic modeling on hydrogen addition and abstraction by H radical of 1,3-cyclopentadiene and the associated chain-branching reactions

Qian Mao[1*], Liming Cai[1], Heinz Pitsch[1]

[1]Institute for Combustion Technology, RWTH Aachen University, Templergraben 64, Aachen 52056, German

*Corresponding author: Qian Mao

Mail Address: Institute for Combustion Technology, RWTH Aachen University, Templergraben 64, Aachen 52056, German

Email: q.mao@itv.rwth-aachen.de

**Abstract**

Cyclopentadiene (CPD) is an important intermediate in the combustion of fuel and formation of aromatics. Specifically, the reaction kinectis of H atom and CPD including abstraction and addition play significant roles in the polycyclic aromatic hydrocarbons (PAH) formation and fuel consumption. In the present study, the kinetics and thermodynamic properties for hydrogen abstraction and addition with CPD, and the related reactions including the isomerization and decomposition on the $C_5H_7$ potential energy surface were systematically investigated by theoretical calculations. High-level ab initio calculations were adopted to obtain the stationary points on the potential energy surfaces of CPD + H. Phenomenological rate coefficients for temperature- and pressure-dependent reactions in the full potential energy surface were calculated by solving the time-dependent multiple-well RRKM/master equation, and the hydrogen abstraction reactions were based on the conventional transition state theory. In terms of the hydrogen abstraction reactions, the hydrogen abstraction from the saturated carbon atom in CPD is found to be the dominant channel. For the hydrogen addition and the associated reactions on the $C_5H_7$ PES, the allylic and vinylic cyclopentenyl radicals and $C_2H_2 + C_3H_5$ were found to be the most important channels and reactivity-promoting products, respectively. Contributions of the addition and abstraction channels were assessed, and the results show that the hydrogen addition reactions are relatively favored over hydrogen abstraction reactions at low temperatures and high pressures. The previously neglected role of open-chain intermediates in the evaluation of the reaction kinetics has been suggested and the corresponding rate constants have been recommended for inclusion in the modeling of the H+ c-C5H6 reaction. Results indicate that the transformation from to straight-chain $C_5H_7$ is kinetically unfavorable due to the high strain energy of the 3-membered ring structure of the isomerization transition state. Moreover, the thermodynamic data and the calculated rate coefficients for both H atom abstraction and addition were incorporated into the ITV Mech to examine the impact of the computed pressure-dependent kinetics of $C_5H_6$ + H reactions on model predictions. Significant improvement in the laminar flame speed of cyclopentadiene was observed, especially in the fuel-rich conditions. In addition, a noticeable improvement was found for the mole fractions of important intermediate species, e.g., acetylene, ethylene, allene, in both of the CPD pyrolysis and oxidation conditions.

# 1. Introduction

The cyclopentadiene (CPD) and its radical cyclopentadienyl (CPDyl) are important intermediates during the pyrolysis and oxidation of conventional hydrocarbons and oxygenated fuels [1–6]. Besides the well-known species such as $C_4H_5$ isomers, $C_3H_3$ to the formation of soot precursors, the CPDyl is also a crucial contributor to the polycyclic aromatic hydrocarbons (PAHs) and soot formation [7–9]. Experimental studies of phenol pyrolysis [4,5] firstly provided mechanistic insights into the role of CPDyl in the formation of aromatics such as naphthalene, indene, which is then confirmed by a series of quantum chemistry calculations [10–14] as recombination of resonance-stabilized radical of CPDyl mainly contribute to the benzene and further growth of PAHs. To understand and model complex combustion systems especially those related to aromatic and soot formation, it is essential to include all pertinent elementary reactions of CPD with precise rate constants.

Recently, some experiments on the combustion and pyrolysis of CPD have been conducted, providing valuable insights into the fuel initiation reactions as well as CPD/CPDyl in PAH formation. According to Ji et al. [15], the measured laminar flame speeds and extinction strain rates of CPD/air mixture are notably determined by the fuel chemistry, and the controlling initiation reaction is the H-abstractions of CPD [15]. Butler and Glassman [16] performed experimental studies of both the pyrolysis and oxidation of CPD at atmospheric pressure near 1150 K in the Princeton plug flow reactor. In the pyrolysis condition, almost 80% of the CPD was pyrolyzed to form naphthalene as a result of CPD/CPDyl reactions. Whereas in the oxidation condition, the concentrations of $C_1$-$C_4$ species increased because of the fuel decomposition. However, in both conditions, the reactions of CPD plus H were crucial for the fuel consumption and aromatics formation [16].

The CPD + H reactions and the associated $C_5H_7$ chemistry are undoubtedly an indispensable part of the chemistry model. In terms of the reaction kinetics of CPD and H, there are quite a few existing experimental data. The global reaction constant for H + CPD reaction was studied by Roy et al. [17] in the shock tube, which however did not include the detailed mechanism of individual rate coefficients, such as abstraction, addition, chemical activation reactions, and requires extensive studies. Initially, the rate constant of hydrogen abstraction reaction of CPD was obtained using the analogous exothermic reactions with formaldehyde [18]. The first detailed reaction mechanism for CPD/CPDyl and the associated $C_5H_7$ isomerization and decomposition reaction was established by Zhang and Bozzelli [19,20] by performing a comprehensive study using the semi-empirical PM3 method, which still acts as the cornerstone of CPD chemistry in most of the kinetics models, such as USC Mech II [21], Aramco Mech 2.0 [22], ITV Mech [23], *etc*. Later, Moskaleva and Lin [24] investigated the potential energy surface (PES) of CPD + H at the G2M level, a relatively more accurate quantum chemistry method than the semi-empirical PM3 method, including the abstraction and addition reactions, whereas only the reaction kinetics of the low-lying product channels was accessed. Previously reported kinetics on $C_5H_7$ PESs remains ambiguous. For example, whether the H addition reaction in forming cyclopentene-3yl (CYC$_5$H$_7$_1), is barrierless, which conflicts in different literatures [19,24]. The chemically activated reaction of $C_5H_6$ + H ↔ A-$C_3H_5$ + $C_2H_2$ is vital for the accurate prediction of fuel consumption and molecular weight growth. Whereas the barrier height for the C-C fission of 1,4-pentadiene-1yl radical (C=CCC=C*) predicted by Zhong and Bozzelli [19] was reported to be 3.5 kcal/mol lower than that from Moskaleva and Lin [24], which would lead to the uncertainty of rate constants by a factor of at least 3-5 at flame temperatures. Besides, only the high pressure limit (HPL) rates of 1,4-pentadiene-1yl radical (C=CCC=C*) decomposition are available in the current chemistry models. According to previous studies on thermal decomposition of $C_3H_7$ [25] and $C_4H_7$ [26], the decomposition of such "weakly-bound free radicals" is far away from the HPL under combustion conditions. Hence the pressure-dependent kinetics on the $C_5H_7$ PES is highly desired.

In the present study, systematic study of the full potential energy surface (PES) of CPD + H system with the emphasis on exploring the addition and abstraction as well as the associated isomerization, decomposition and chemical activation reactions was explored. High pressure limit rate constants of the hydrogen abstraction reactions by hydrogen atom was computed using the conventional transition state theory, while the time-dependent, multiple-well master equation (RRKM/ME) was employed to derive the temperature- and pressure-dependent rate constants for the hydrogen addition reactions as well as the subsequent isomerization, decomposition reactions. The calculated rate constants were then fitted to functions in Arrhenius form and incorporated into the ITV Mech [23,27] to examine the effects of pressure-dependent kinetics on model predictions of CPD pyrolysis and combustion as well as aromatics formation from various fuels in a wide range of operating conditions.

## 2. Methodology

### 2.1 Theoretical calculation

Reactions of hydrogen abstraction/addition of CPD, isomerization between the cyclo- and

straight-$C_5H_7$, and the decomposition channels were fully explored using electronic structure calculations. The optimized geometries of all stationary points of reactants, transition states, wells and products on the PES of CPD + H were obtained using density functional theory (DFT) with the hybrid exchange-correlation functional M06-2X [28] and the 6-311+G(d,p) basis set. The vibrational frequencies and zero-point energies (ZPEs) were calculated at the same level and scaled by a factor of 0.97 [29,30]. Restricted and unrestricted DFT were employed to closed-shell molecules and open-shell species, respectively. The transition state was confirmed to have exactly one imaginary frequency and connect to the corresponding reactants and products on both sides by the intrinsic reaction coordinate (IRC) analysis at the same level of structure optimization and frequency analysis. The multi-reference characteristic parameter, $T_1$ diagnostics, were examined for all the species involved in this study by CCSD(T)/cc-pVTZ calculations and found to be smaller than the thresholds (for closed-shell species is 0.02 and the open-shell species is 0.045) [31,32] as listed in the Supplementary Material. This indicated that the reactions considered in the current study were weakly correlated and the single-reference method was reliable to obtain energies of reactions and barrier heights. Single point energies (SPEs) were calculated at the CCSD(T)/cc-pVDZ and CCSD(T)/cc-pVTZ levels. Then the energies were extrapolated to the complete basis set (CBS) limit according to the two-point scheme proposed Truhlar [33]:

$$E_{CBS} = \frac{3^\alpha}{3^\alpha - 2^\alpha} E_{X=3}^{HF} - \frac{2^\alpha}{3^\alpha - 2^\alpha} E_{X=2}^{HF} + \frac{3^\beta}{3^\beta - 2^\beta} E_{X=3}^{cor} - \frac{2^\beta}{3^\beta - 2^\beta} E_{X=2}^{cor} \quad (1)$$

where $E^{HF}$ and $E^{cor}$ are the Hartree-Fock (HF) and the correlation energies; For CCSD(T) method, exponent parameters of $\alpha$ and $\beta$ equal 3.4 and 2.4; X = 2 and 3 represent the basis-sets of cc-pVDZ and cc-pVTZ, respectively. All the quantum chemistry calculations were performed by the Gaussian 09 program package [34].

For reaction channels with a well-defined transition state, as the hydrogen abstraction from CPD, high-pressure coefficients were obtained based on conventional transition state theory (TST) through employing the rigid-rotor harmonic-oscillator (RRHO) approximation for all degrees of freedom of CPD and transition states. For the hydrogen addition and other associated reactions on the CPD+H PES, the temperature and pressure dependent rate coefficients were calculated by Rice–Ramsperger–Kassel–Marcus/Master Equation (RRKM/ME) with the temperature ranging from 800 to 2500 K and pressure from 0.01 to 100 atm. The low-frequency modes of the internal rotors such as methyl and ethyl like torsions of species were treated as 1-dimensional hindered rotors [35,36] with hindrance potentials computed at the level of M06-2X/6-311+G(d,p) same with the geometry optimization and frequency analysis. For other vibrational modes, harmonic oscillator assumptions were employed to evaluate the densities of states for stationary points. The asymmetric Eckart tunneling model [37] was taken into account for tunneling correction, which provides similar results to small-curvature tunneling (SCT) approximation above 500 K [38]. The 'single-parameter exponential down' model [39,40] expressed by $<\Delta E_{down}> = 400 \times (T/300)^{0.7}$ cm$^{-1}$ was used as the energy transfer model in MEs, which has been widely used in the study of CPD/CPDyl and aromatics [9,41]. The L–J parameters for the interaction between the reactants and the bath gas Ar were the same with the previous study of CPD + H system by Moskaleva and Lin [24], for Ar, $\sigma$ = 3.47 Å, $\varepsilon$ = 79.23 cm$^{-1}$ ; for CPD, $\sigma$ = 5.78 Å, $\varepsilon$ = 273.8 cm$^{-1}$. The phenomenological rate coefficients for reactions involved in the PES of hydrogen addition and associated reactions were computed by using the MESS code [42] through solving the MEs based on chemically significant eigenstate approach proposed by Miller and Klippenstein [43–45]. All the temperature and pressure dependent rate coefficients of the investigated reactions were fitted into the single Arrhenius equations as a function of temperature,

$$k = AT^n \exp(-E_a / RT) \qquad (2)$$

where $A$ is the pre-exponential factor in the unit of cm$^3$mole$^{-1}$s$^{-1}$ or s$^{-1}$, $T$ is the temperature in units of Kelvin, $n$ is the temperature exponent, $E_a$ is the activation energy in the unit of cal/mol. The rate coefficients of the reactions of kinetic importance at different pressures were fitted in "PLOG" formats as shown in the Supplementary Materials, which can be readily used in the hierarchical development of reaction mechanisms for CPD pyrolysis and oxidation.

**2.2 Thermodynamic data**

The thermodynamic data (formation enthalpy, entropy and heat capacity) of species such as $C_5H_6$, $C_5H_5$ and $C_5H_7$ isomers shown in the PES of CPD + H were determined using the calculated energies, frequencies and barriers at the CCSD(T)/CBS//M06-2X/6-311+G(d,p) level of theory. The standard-state of heat formation, $\Delta_f H^0_{298K}$, was computed from the atomization energy and the experimental heats of formation of the atoms in Eq. 3,

$$\Delta_f H^0_{298K} = \sum_{j=H,C,O} x_j \left( \Delta_f^{exp} H^0_{0K,j} - \left[ H^0_j(298K) - H^0_j(0K) \right] \right) - \sum D_0 + \left[ H(298K) - H(0K) \right] \quad (3)$$

where $x_j$ is the number of atom type $j$ in the molecule, $\Delta_f^{exp} H^0_{0K,j}$ is the experimental formation enthalpy of atoms at zero Kelvin from NIST- JANAF [46]. $\sum D_0$ is the atomization energy of the molecule expressed as,

$$\sum D_0 = \sum_{j=H,C,O} x_j E_{0,j} - E_0 - E_{ZPE} \quad (4)$$

where $E_{0,j}$ is the calculated total electronic energy of the atom type $j$ including spin-orbit corrections, $E_0$ is the CCSD(T)/CBS electronic energy of the molecule, and $E_{ZPE}$ is the M06-2X/6-311+G(d,p) zero-point energy scaled by 0.97 [29,30]. In addition, the thermochemical values based on the additivity methods proposed by Benson [47] are summaried in Table 1 for comparsion. Difference between the two sets of enthalpy values are smaller than 2 kcal/mol for most of the species, which are within the estimated uncertainty limit of the computational enthalpy values [48]. However, the difference in enthalpy increases to as large as about 3 kcal for branched-$C_5H_7$ and even 5 kcal/mol for RSR 1,3-Pentadien-5-yl. Similarly, the entropy values determined using computational methods are similar to those calculated using THERM, with differences ranging between 0.0 and 3 cal/mol/K, except for W1, allylic cyclopentenyl, which is as large as 6.5 cal/mol/K. W1, W2 and W3 are all five-member rings and their entropy should be similar as well as relative higher than $C_5H_5$ and $C_5H_6$. However, results from THERM based on group additivity indicates the entropy of $C_5H_6$ and W1 are similar and much lower than the other two cyclopentenyl radicals. The thermochemical parameters were calculated in 100 K increments from 300 to 3000 K, and the results were then fitted to the standard 14 parameter NASA polynomial which were detailed in the Supplementary material.

Table 1 Thermochemical values of some key intermediates lying in the PES of CPD+H ($\Delta H_{298}$ in the unit of kcal/mol, and $S_{298}$, $C_p$ in the unit of cal/mol/K)

|  | $\Delta H_{298}$ | | $S_{298}$ | |
| --- | --- | --- | --- | --- |
|  | Present work | THERM | Present work | THERM |
| $C_5H_6$ | 32.67 | 32.60 | 66.63 | 65.50 |
| $C_5H_5$ | 64.50 | 63.60 | 68.27 | 66.79 |
| W1 (allylic cyclopentenyl) | 40.88 | 38.90 | 72.15 | 65.60 |
| W2 (alkylic cyclopentenyl) | 53.11 | 53.30 | 71.14 | 73.92 |
| W3 (vinylic cyclopentenyl) | 70.11 | 70.30 | 70.35 | 72.60 |
| W4 (1,3-Pentadien-5-yl) | 53.94 | 48.90 | 74.51 | 75.82 |
| W5 (1-Pentyn-5-yl) | 84.23 | 86.30 | 78.93 | 82.41 |
| W6 (1,4-Pentadien-1-yl) | 85.11 | 84.50 | 76.32 | 82.75 |
| W7 (1,2-Pentadien-5-yl) | 82.90 | 83.09 | 80.03 | 83.02 |
| W8 | 63.56 | 62.55 | 71.79 | 69.01 |
| W9 | 57.18 | 54.08 | 75.26 | 73.47 |
| W10 | 81.62 | 78.08 | 77.25 | 76.65 |

## 2.3 Kinetic modeling

The computational results including both the kinetics and thermodynamics data of the initiation reactions of CPD were incorporated into the ITV mechanism, which contains sub-mechanisms for base, $C_5$, and PAH chemistry. A detailed description of the ITV mechanism can be found in Ref. [23,27]. For convenience, this base and updated mechanisms were referred to as 'ITV Mech' and 'ITV Mech_updated' hereafter, respectively. For reactions with rate constants greater than 100 cm$^3$ mol$^{-1}$s$^{-1}$ or s$^{-1}$ at 750 K, the threshold proposed by Rashidi et al. [49], were included in the mechanism. Therefore, in total, 14 reactions (hydrogen addition and abstraction of CPD, isomerization among $C_5H_7$ isomers, unimolecular decomposition of $C_5H_7$ isomers and chemically activated reactions) were updated or added in the ITV Mech as listed in the Supplementary Materials.

To reveal the impact of the reactions involved on the CPD + H potential energy surface (abstraction, addition, isomerization and decomposition reactions) on CPD combustion, the laminar flame speed of CPD and species mole fraction profiles in both the pyrolysis and oxidation conditions were simulated and compared with experimental data. The laminar flame speed of cyclopentadiene was determined by Ji et al. [15] in a counterflow configuration under ambient pressure with equivalent ratio from 0.7 to 1.5. Numerical simulations were performed using the in-house code FlameMaster version 4.1.0 [50]. The effect of the gridpoint on the accuracy of the results was evaluated and was found that a gridpoint of 800 was enough to avoid the discretization error and ensure a good accuracy of the calculations. The experiments of both the pyrolysis and oxidation of cyclopentadiene were studied by Butler and Glassman [16] in the Princeton's turbulent adiabatic plug flow reactor at 1 atm with varied equivalent ratios and initial temperatures. In the present study, the pyrolysis condition with $\phi$ = 97.8 at $T_0$ = 1202 K and oxidation condition with $\phi$=1.03 at $T_0$ = 1198 K were simulated and the time evolution of the species mole fraction profiles predicted using both the ITV Mech and ITV Mech updated were compared against the experimental data. Besides acting as a fuel in the combustion, cyclopentadiene is also an important intermedium species in the combustion of large hydrocarbons. Here the combustion of n-heptane in the counterflow burner at 1 atm was compared against key intermediate species from mass spectrometric studies.

# 3. Pressure-dependent reaction kinetics

## 3.1 Potential energy surface

The molecular structure of the CPD as well as the C-H bond dissociation energies (BDE) of CPD at 298.15 K calculated at the CCSD(T)/CBS//M06-2X/6-311+G(d, p) level of theory are shown in Figure 1. The CPD molecule has a symmetry number of two and three different types of C-H bonding. It is noted that the BDE of the allyl C-H bond as shown of site 1 is 84.8 kcal/mol in the present study, which agrees well with the BDE determined by the experimental study with time-resolved photoacoustic calorimetry at 85.3 ±1.7 at 298.15 K [51]. The BDEs of the vinly C-H bond at the other two sites are similar and about 23-24 kcal/mol higher than that at site 1. Such huge difference in the BDEs of the C-H bond at allyl and vinly is also found in cyclopentene [52].

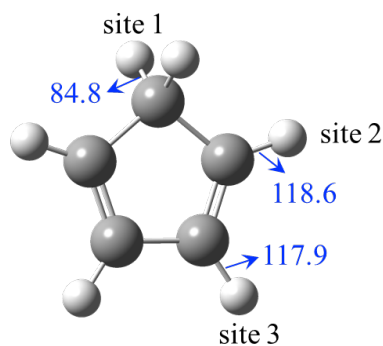

Figure 1. Molecular structure of CPD and the BDE of C-H bonds calculated in the present work (unit, kcal/mol). The label $i$ ($i$ = 1, 2, 3) denotes the C-H site in the CPD molecule

The results of the electronic structure calculations for the $C_5H_6$ + H PES including the abstraction (blue), addition(red), isomerization(purple) and dissociation channels are displayed in Fig. 2 at CCSD(T)/CBS//M06-2X/6-311+G(d, p) level of theory. Cartesian coordinates of all reactants, transition states, wells and products in Fig. 2 are provided in the Supplementary material. Firstly, for the hydrogen abstraction reactions, it is preferred to take place at site 1, as its barrier height is 5.2 kcal/mol and about 13 kcal/mol higher than that at site 2 and site 3. For the addition reaction, H atoms can add to the double bond of CPD either at site 2 to form allylic cyclopentenyl (W1) at a barrier height of 2.4 kcal/mol or site 3 to form alkylic cyclopentenyl (W2) at a barrier height of 4.1 kcal/mol. The energy of W1 is - 43.6 kcal/mol relative to the reactant ($C_5H_6$ + H) and has the lowest energy on the PES. The barrier height of the hydrogen addition at site 3 is just 1.1 kcal/mol lower than the hydrogen abstraction at site 1, indicating the abstraction and addition channels are competitive under combustion conditions. The isomerization between cyclopentenyl W1 and W2 is thermally accessible as the barrier height is 48.3 kcal/mol, corresponding to 4.7 kcal/mol above the reactant ($C_5H_6$ and H), while the barrier height of W1 isomerizes to vinylic cyclopentenyl (W3) by hydrogen transfer is 76.2 kcal/mol, which is thermally unfavorable compared to the isomerization to form W2. The isomerization between W2 and W3 is not included in the PES as it the isomerization is indirect. That is, W2 firstly form the intermediate specie of cycC4cycC3, which, however, needs to overcome a tight three-membered ring transition structures, which has high strain energy. The above three cyclopentenyls (W1, W2 and W3) can experience the ring-opening through C-C $\beta$-scission by forming the straight-chain $C_5H_7$ radicals (W4, W5, W6, W7). The barrier height of W1 to 1,4-pentadiene-3yl (W4) and W2 to 1,4-pentadien-1-yl (W6) are about 45 kcal/mol, which are about 15 kcal/mol higher than the homolytic C-C cleavage of W3 to 1-pentyn-5-yl (W5) or 1,2-pentadien-5-yl (W7). Therefore, the isomerization reactions between allylic (W1) and alkylic (W2) cyclopentenyl are competitive to their dissociation channels, while that of vinylic cyclopentenyl (W3) favors the dissociation channels than its isomerization. Compared to the other three straight-chain $C_5H_7$ radicals, W4, a resonance stabilized radical, has the lowest energy. The isomerization among the above four straight-chain radicals are significantly higher than the dissociation channels by C-C $\beta$-scission as indicated in the PES. Here, only the isomerizations among the four straight-chain radicals directly generated from cyclopentenyl radicals are considered in the PES as the barrier height of their isomerization to the rest pentadieneyl and pentynyl radicals by hydrogen transfer are much tight (at the energy of at least 30 kcal/mol above the reactant) than their dissociation channels. For W5, W6, and W7, the dissociation channel includes both C-H and C-C $\beta$-scissions except for W4 that only has the C-H $\beta$-scission. The barrier heights of the C-H $\beta$-scissions of W5, W6, W7 are 15.4, 13.1 and 9.9 kcal/mol higher their C-C $\beta$-scission channels, respectively. Specifically, the most feasible path for W6 is to $C_2H_2$ and allyl (A-$C_3H_5$), whose energy lies in the bottom of the products in the addition channel. The low energy exit channels for W5 and W7 lead to the product of $C_2H_4$ and propargyl ($C_3H_3$), which are 9.96 kcal/mol above the reactant. The above exit channel is comparable to the reaction pathway to form P1 ($C_2H_2$ and allyl), which, however, is missing in the previous studies of the $C_5H_6$ +H reactions [19,24] and kinetics model of cyclopentadiene chemistry.

Despite the isomerization between cyclopentenyl radicals and straight-chain $C_5H_7$ radicals, the formation of branched $C_5H_7$ radicals is also favored as indicated in the PES. The isomerization paths from straight-chain 1,2-pentadien-5-yl (W7) to branced isopentadiene radical (W9) involve W7-TS78-W8-TS89-W9 through passing a four-membered cyclic backbone transition state (TS78). The highest bottleneck in the above isomerization paths is 13.5 kcal/mol, much lower than the dissociation channel of W7, which are at energies of 23.3 kcal/mol (C-C $\beta$-scission) and 33.2 kcal/mol (C-H $\beta$-scission). It indicates that the isomerization path from straight-chain W7 to branched $C_5H_7$ W9 should play a significant role comparing to its consumption routes. The above isomerization from straight-chain $C_5H_7$ to branched $C_5H_7$ is relatively easier than the isomerization from the straight-chain $C_4H_7$ to branched $C_4H_7$ as the transition state with the three-membered rings has higher strain energy than the four-membered rings [26]. The exit channel for W9 by C-C $\beta$-scission to form vinyl ($C_2H_3$) and allene (A-$C_3H_4$) is at the similar energy to its isomerization to isoprene radical (W10), indicating the above two routes compete with each other. Compared to W9, the exit dissociation channels of W10 requires low energy barriers. Specifically, the lowest energy route leads to methyl ($CH_3$) and vinylacetylene ($C_4H_4$), and the other is vinyl and propyne (P-$C_3H_4$). The energy of propyne is found to be 0.8 kcal/mol lower than that of allene in the present study, which is 0.9 kcal/mol from QCISD(T)/ cc-pvtz calculations by Miller and Klippenstein [44].

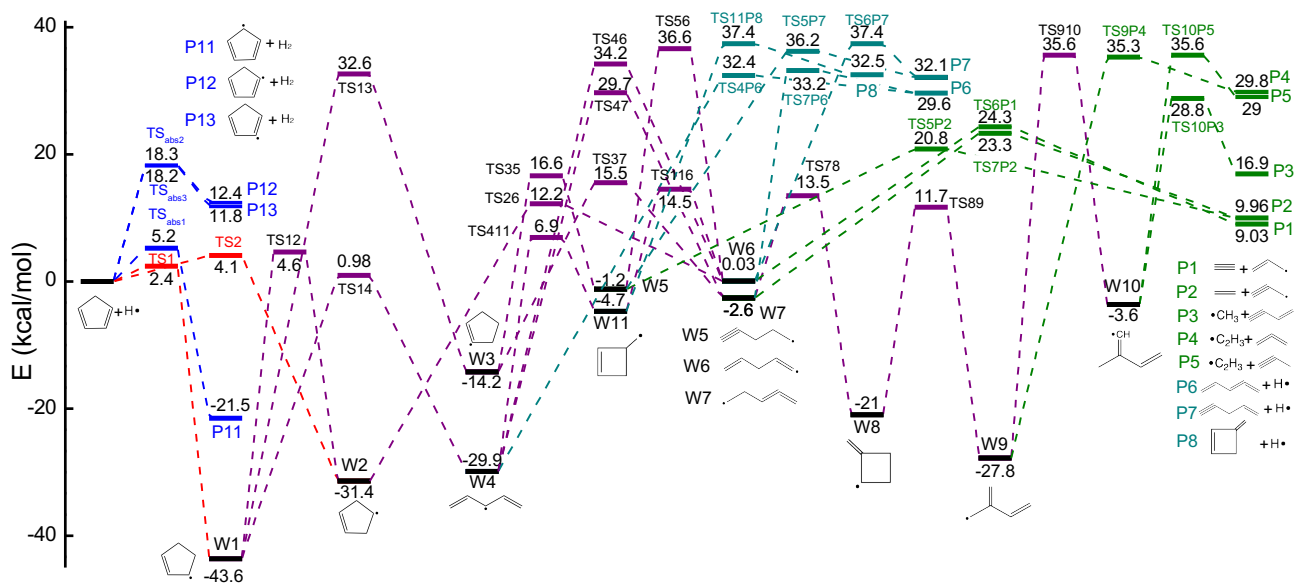

Figure 2. Potential energy surface (PES) of CPD + H at the CCSD(T)-CBS//M06-2X/6-311+G(d,p) level of theory (unit, kcal/mol). The abstraction channels highlighted in blue, the addition channels in red, the isomerization channels in purple, the dissociation channels of C-H fission and C-C fission colored in cyan and green, respectively.

Energies of some stationary points on the PES are selected and summarized in Table 2 by comparing with that from DFT method of M06-2X/6-311+G(d, p) together with previous theoretical results based on the modified Gaussian-2 (G2M(RCC, MP2)) method from Moskaleva and Lin [24], CBS-QB3 level of theory from Sharma et al. [53], semi-empirical method of PM3 by Zhong and Bozzelli [19]. Energies calculated based on the CCSD(T)/CBS//M06-2X/6-311+G(d,p) are generally regarded to be accurate with a derivation less than 1 kcal/mol from experiments [33,54]. Compared to energies calculated from the CCSD(T)/CBS//M06-2X/6-311+G(d, p), the DFT method of M06-2X/6-311+G(d, p) has a maximum deviation of less than 1 kcal/mol for most of the species, which agrees well with the results of hydrogen abstraction from benzene [54]. The H addition at site 2 was reported to be barrierless at both the G2M level [24] and CBS-QB3 level of theory [53], and 2.70 kcal/mol by PM3 method [19]. The results based on the G2M method [24] are overestimated by about 3 kcal/mol compared to that from the CCSD(T)/CBS, which is similar to that from CBS-QB3 method[53]. The semi-empirical method of PM3 has a deviation of about 5 kcal/mol except for TS14, the transition state of the W1 and W4, by about 25 kcal/mol. Such a huge difference in the barrier height could give rise to severe uncertainties in the rate constant. However, the reaction kinetics of $C_5H_6 + H$ based on the PM3 level of theory is currently the most widely used in the chemical kinetics models, such as USC Mech II [21], Aramco Mech 2.0 [22], ITV Mech [23].

Table 2 Energies of selected species on the PES of $C_5H_6 + H$ (unit: kcal/mol)

| Species | CCSD(T)/CBS//M06-2X/6-311+G(d,p) | M06-2X/6-311+G(d,p) | G2M(RCC,MP2) [24] | CBS-QB3[53] | PM3[19] |
|---|---|---|---|---|---|
| Wells | | | | | |
| W1 | -43.58 | -42.93 | -40.22 | -43.18 | -43.47 |
| W2 | -31.37 | -31.29 | -28.72 | — | -29.19 |
| W4 | -29.85 | -29.53 | -26.81 | -36.55 | -34.05 |
| W6 | -0.26 | 0.03 | 1.38 | — | 1.15 |
| W11 | -4.67 | -4.83 | -3.41 | — | — |
| Transition States | | | | | |
| $TS_{abs1}$ | 5.23 | 6.11 | 6.34 | | — |
| TS1 | 2.40 | 2.14 | 0 | 0 | 2.70 |
| TS2 | 4.07 | 4.43 | 4.32 | — | 2.70 |
| TS12 | 4.65 | 4.08 | 7.42 | — | — |
| TS14 | 0.98 | 2.38 | 3.62 | -0.34 | -24.25 |
| TS26 | 12.20 | 13.84 | 14.68 | — | 11.15 |
| TS116 | 14.45 | 16.06 | 16.96 | — | — |
| TS6P1 | 24.29 | 25.41 | 27.37 | — | 18.95 |
| TS6P7 | 37.37 | 37.59 | 37.98 | — | — |
| Products | | | | | |
| P1 | 9.05 | 11.56 | 10.68 | | 11.25 |
| P2 | 9.96 | 12.11 | — | 11.1 | — |
| P7 | 32.50 | 32.57 | 32.28 | | |
| P11 | -21.45 | -21.37 | -21.01 | | |

## 3.2 Reaction kinetics of hydrogen abstraction reaction of cyclopentadiene

The hydrogen abstraction reaction of the cyclopentadiene by hydrogen atom is calculated based on the conventional transition state theory. The high-pressure limit (HPL) rate constants of the hydrogen abstraction reactions at three different sites of the CPD are illustrated in Figure 3. The rate constants for hydrogen abstraction at site 2 and site 3 are quite similar as indicated in Fig. 3(b). Difference in the barrier height of the hydrogen abstraction leads to great difference in the reaction rate particularly at low temperature regions. It is approximately three orders of magnitude larger for the reaction at site 1 than that at the other two sites at 800 K, and the difference decreases with the increasing temperature as it comes to two times at 2500 K. In kinetics models, only the reaction at site 1 is considered. The rate constants of the above three hydrogen abstraction reactions from the the rate rule are also included in Fig. 3. It overpredictes the hydrogen abstraction rate constants at site 1 by about five times, while underpredicts that at site 2 and 3 compared to our present results.

Theoretical studies the hydrogen abstraction from CPD mainly focuses on the reaction at site 1. Therefore, the rate constant of hydrogen abstraction at site 1 from the present study is compared with those from experiments [17,55] and theoretical studies [19,20,24,56,57] as shown in Fig. 3 (a). The hydrogen abstraction reaction is one of the most important and sensitive reactions of fuel consumption and aromatics formation. However, large discrepancies of up to a few orders of magnitude exist among the reported rate constants and kinetics models. Specifically, Roy et al. [17,55] estimated the rate constant by fitting to the experimental data of CPD oxidation. This rate constants was usd in ITV Mech[23]. Emdee et al. [56] obtained the reaction rate based on the analogous exothermic reactions with formaldehyde [18]. Zhong and Bozzelli [19] applied the semi-empirical PM3 method to derive the hydrogen abstraction rate and it was used in Aramco 3.0 [22]. Moskaleva and Lin [24] calculated the HPL rate constant from transition state theory with energies at the G2M(RCC, MP2) level, which was adopted in the USC Mech II [21]. Since the barrier height from our current study is 0.8 kcal/mol lower than that from Moskaleve and Lin [24], it leads to a difference in the reaction rate by a factor of 1.5 at 2500 K. Djokic et al. applied analogy rules to derive the rate constant of the hydrogen abstraction reaction, which is used in the CRECK Mech [57]. It should be noted that the reaction rate from our present study lies in between the reported rate constants.

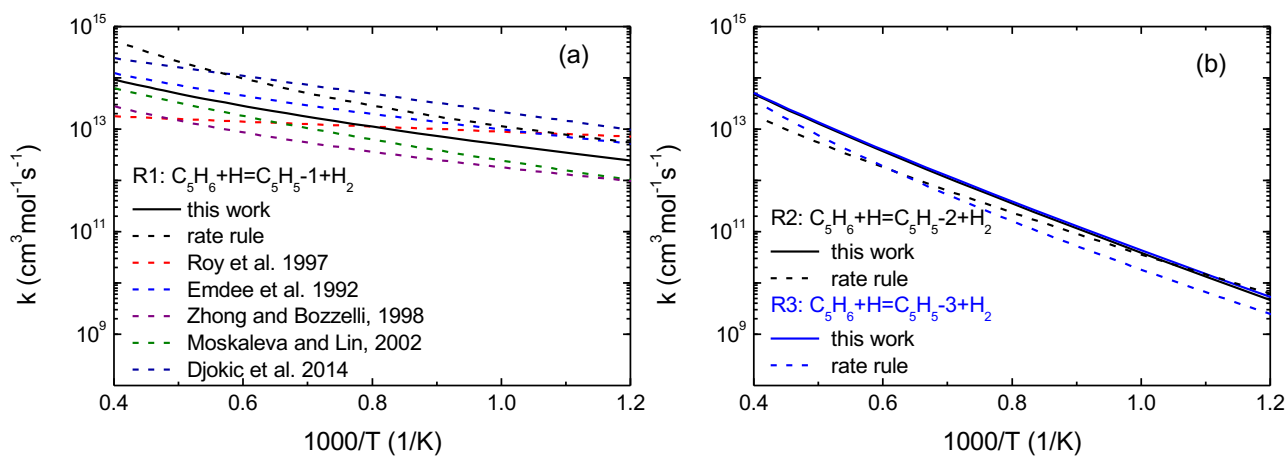

Figure 3. High pressure limit rate constants for hydrogen abstraction reactions (solid lines are results from the present study, dash lines are the hydrogen abstraction rate constants in literatures)

### 3.3 Hydrogen addition reaction of cyclopentadiene

The hydrogen addition reaction at sites 2 and 3 by forming allylic cyclopentenyl (W1) and alkylic cyclopentenyl (W2) are important entrance channels on the $C_5H_6$ + H PES. The rate constants of the above addition reactions are illustrated in Fig. 4 (a) and (b), respectively, together with the HPL rate constants from literatures [19,58]. Similarly, the rate constants of the above two addition reactions vary with temperature and pressure and obvious fall-off effect is observed at high temperatures. The derivation of the rate constants to the HPL rate constants increases with the increasing temperature as well as decreasing pressure. Taking the hydrogen addition reaction R4 for example, at 1000 K, the rate constant at 0.01 atm is three times smaller than the HPL rate constants, whereas the difference increases to two orders of magnitude at 1500 K. Besides, the fall-off effect takes place above 1500 K at 100 atm, and it decreases to the temperature lower than 800 K at 0.01 atm. Therefore, the rate constants of H addition to CPD under typical combustion conditions are probably far from their high pressure limit. This emphasizes the necessity of using accurate pressure-dependent rate coefficients in combustion mechanisms. Moreover, the rate constant presents a negative trend and as pressure increases its rate constant gradually becomes negligible. This phenomenon originates from competition between collision and chemical reaction. The allylic and alkylic cyclopentenyl radicals are thermally unstable at high temperatures. This consequently leads to a negative temperature dependence of the rate constant at high temperatures, which is enhanced with the decreasing pressure. The HPL rate constants of the above two hydrogen addition reactions are overpredicted based on the RGM rate rule [58] by a factor of 3-5. The rate constant for hydrogen addition to CPD at site 2 was about one half of that at site 3 provided by Zhong and Bozzelli [19] based on PM3 calculations as the barrier heights of the H addition at both sites were reported to be the same, which however has a difference of 1.7 kcal/mol in our current study. This consequently underestimates the rate constant for hydrogen addition at site 2 for the temperature of interest and overestimated that at site 3 below 2000 K.

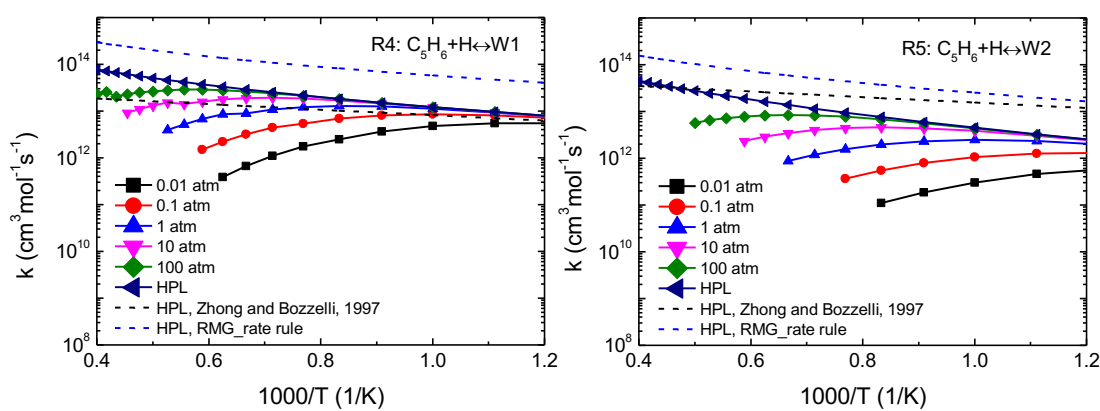

Figure 4. Pressure dependent rate constants of hydrogen addition reactions: R4: $C_5H_6+H\leftrightarrow W1$ and R5: $C_5H_6+H\leftrightarrow W2$. The missing rate coefficients at low pressure high temperature are due to well-merging

The formation of the RSR W4 from reactant of $C_5H_6$ + H through "well-skipping" pathways is the most energetically favorable and kinetically significant among other wells on the $C_5H_6$ + H PES. Figure 5 shows the rate constants of the well-skipping reaction of R6: $C_5H_6$ + H ↔ W4 at pressure from 0.01 to 100 atm. The reaction of $C_5H_6$ + H to branched RSR W9 is also included in Fig. 5 as it is the most stable branched $C_5H_7$ radical on the PES. W9 is proceeded over multiple transition states from reactants, and the transtiton state of TS13 at an energy of 32.6 kcal/mol is the bottleneck of the above pathway. This directly leads to the differec in the reaction constant of R6 and R7 by two orders of magnitude. The rate constants of the above two well-skipping reactions present a complex pressure and temperature dependence. The rate coefficient at high pressures overcomes that at low pressures at the high temperature region, while it comes contrary at the low temperature region resulting from the multi-well property of the reaction. Take $C_5H_6$ + H ↔ W4 for example, at 0.01 atm the reaction constant is quickest below 900 K and it goes contrary at high temperatures. In addition, the rate constants of $C_5H_6$+H ↔ W4 from Moskaleva and Lin [24] at 1 and 20 atm are also shown in Figure 5, which present a similar temperature and pressure dependence. However, the HPL rate constant of reaction R6 provided by Zhong and Bozzelli [19] is greatly overpredicted in the temperature range of interested. The rate constants for the formation of W3, W5, W6, W7, W10 as well as the four-membered radicals of W8 and W11 are not reported as these well are chemically unstable at fairly low temperature and their rate constants are also quite low, which are not the focus of the current study.

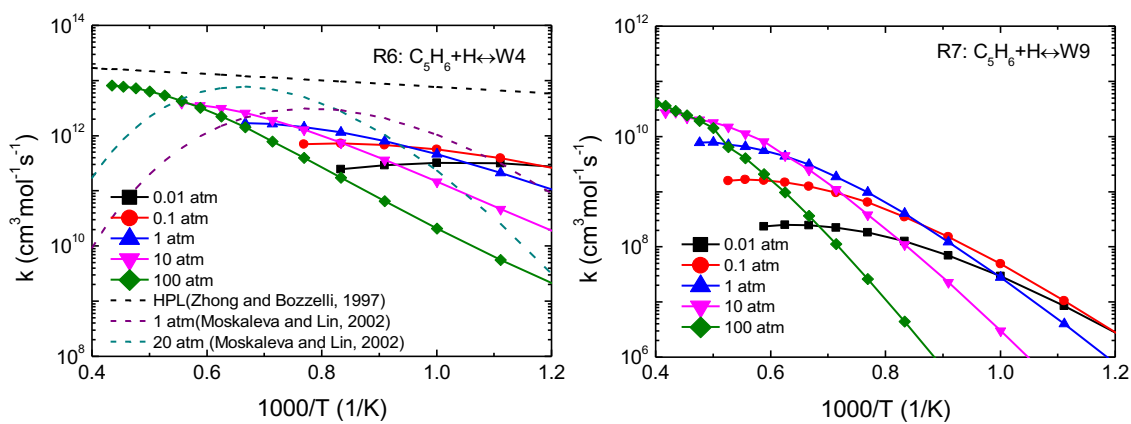

Figure 5. Pressure dependent rate constants of 'well-skipping' reaction of $C_5H_6+H \leftrightarrow W4$ and $C_5H_6+H \leftrightarrow W9$. The missing rate coefficients at low pressure high temperature are due to well-merging

The decomposition pathwtay of $C_5H_6 + H$ to biomolecular products (P1-P5) by "well-skipping" reactions are significantly important for the fuel consumpation and their product are sentensive to aromatic formation. Therefore, precise reaction kinetics of the above two reactions are indispensable for kinetics model prediction. The pressure and temperature dependent rate constants of the above five reactions give similar temperature and pressure dependence. Therefore, only two of the most largest rate constants of the formation of P1($C_2H_2$+A-$C_3H_5$) and P2($C_2H_4$+$C_3H_3$) are considered and plotted in Figure 6, and the others are detailed in the Supplementary Materials. Generally, the rate constants present a negative pressure dependence, which reach their low pressure limit above 2000 K. Rate constants of the formation of P1 and P2 are comparable, while in most of the kinetics model, such as USC Mech II [21], Aramco Mech 2.0 [22], ITV Mech [23], only the formation of biomolecular P1 is considered for the $C_5H_6 + H$ reaction. Moreover, in previous quantum chemistry calculates [19,24], only P1 is considered in the PES of $C_5H_6 + H$. Furthermore, the product of $C_3H_3$ from the reaction of $C_5H_6 + H \leftrightarrow$ P2 is a key species for aramotics formation. Compared to the rate constants for R8 provided in the literatures [19,23,57], it is found our results lies in between especially at high temperature regions. Specifically, Zhong and Bozzelli [19] and Narayanaswamy et al. [23] overpredicted the rate constants, while Djokic et al. [57] underpredicted the rate constants at high temperatures but overpredicted at low temperatures. For the formation of P2 from R9, the estimated rate constant [59] is added for comparison, which is larger than the rate constant obtained from current study. The derivation decreases with the increasing temperature as it is roughly one order of magnitude larger at 2500 K and five orders of magnitude at 800 K, 0.01 atm.

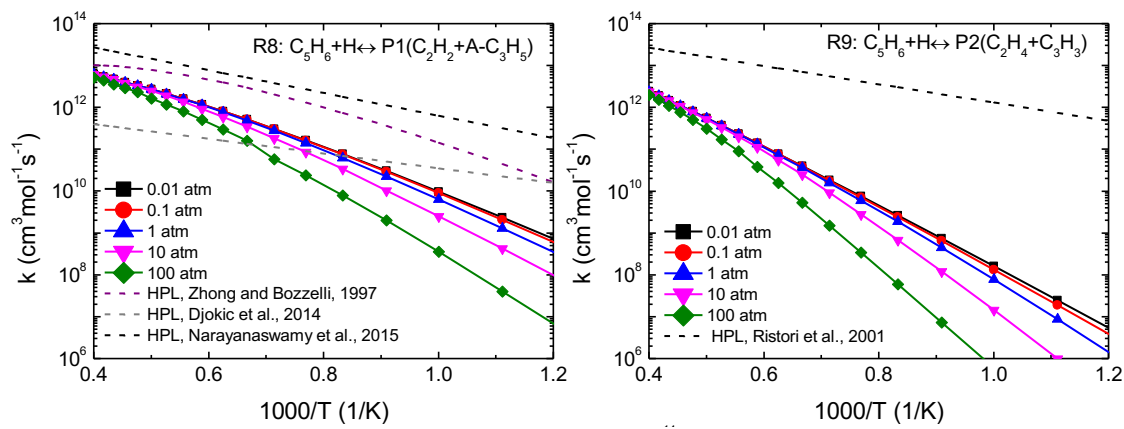

Figure 6  Pressure dependent rate constants for $C_5H_6$ + H to biomolecular products of P1 and P2

The reactions of R4, R5, R6, R7, R8 and R9 account for above 99% of the total reaction rate of hydrogen addition channel of $C_5H_6$ in the entire ranges of pressure and temperature of interest, thererfore the rest reactions consequently can be neglected. The branching ratios of the above six reactions are presented in Figure 7 as a function of temperature at (a) 0.01 atm, (b) 1.0 atm and (c) 100 atm. In general, the dominating reaction varies with the pressure and temperature. Obviously, the total rate coefficient of CPD + H is dominantly determined by the direct hydrogen addition channels by forming cyclopentenyl at low temperatures and that goes contrary at high temperatures as the reaction pathway of the formation of biomolecules become important. Specifically, the branching ratios of the H addition reactions of R4 and R5 decrease significantly with the increasing temperature, while its decreasing rate decreases with the increasing of pressure. The addition channels are still the most important pathway for C5H6+H reaction at high temperatures at 100 atm compared to that at 0.1 at, ,

indicating the addition channel is favored at low temperatures and high pressures. On the contrary, the branching ratios of the formation of W4 and W9 by "well-skipping" reactions increase with the increasing pressure and temperatures. the addition reaction R4 and R5 dominate at low temperatures, as their branching ratios decrease significantly with the increasing temperature. For the well-skipping reaction of R6, its branching ratio firstly increases with the increasing temperature and then decreases, which is preferred to be important at high temperatures and pressures. However, for the reactions of R8 and R9, the branching ratios increase with the increasing temperature and become more important at high temperatures. In general, the reaction of CPD with H radical contributes to molecular growth at high pressures and tends to produce bimolecular products that promote chain branching at low pressures

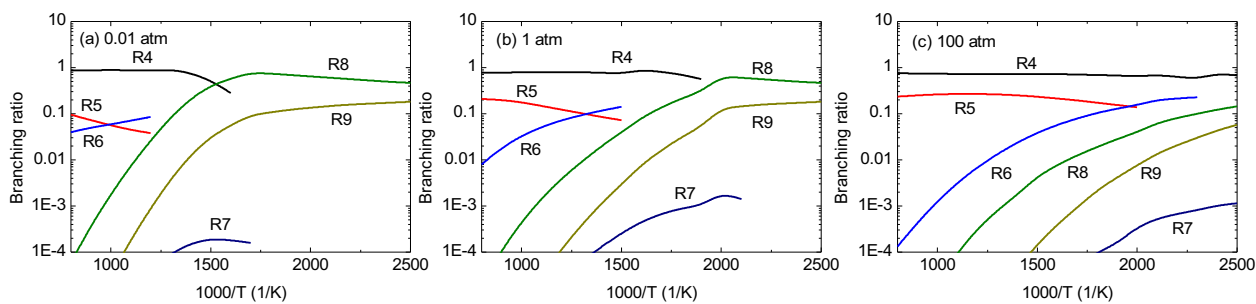

Figure 7 Branching ratios of the of $C_5H_6$+H reactions at pressures of 0.01 atm, 1 atm and 100 atm. (R4: CPD+H↔ W1, R5: CPD+H↔ W2, R6: CPD+H↔W4, R7: CPD+H↔W9, R8: CPD+H↔ P1(A-$C_3H_5$+$C_2H_2$), R9: CPD+H↔ P2($C_3H_3$+$C_2H_4$))

The reaction kinetics of CPD and H was studied by Roy et al. [17] by performing the pyrolysis and oxidation reaction of CPD in the shock tube at pressures between 0.7 and 5.6 bar. Since the results from experiments are not product specific, the rate constants of CPD + H includes all the available channels at the pressure of 0.1, 1 and 10 atm from our current study are shown in Figure 4. Besides, the experimental data are presented in circles [55] and squares [17] as well as the result from previous theoretical calculation at the G2M(RCC, MP2) level of theory at 2 bar [24]. The experimental data is found to be around the predicted rate constant at 0.1 atm from our calculation, The predicted rate constants at 0.1 atm and 10 atm differ by a factor of 2-3. According to Roy et al. [55] that an improved calibration of H-absorption profile yielded the change of reaction rate by a factor of 2.8. Therefore, the predicted reaction rate of $C_5H_6$ + H is within the experimental uncertainties. Moreover, the individual rate constants from the abstraction (R1, R2 and R3) and addition channels at 1 atm are plotted respectively. It is notably that the $C_5H_6$ + H reaction is govered by the abstraction reactions at high temperatures, whereas the addition channel becomes important at low temperatures. Rate constants calculated by Moskaleva and Lin [24] at 2 bar is much larger than the experiment data especially at temperature lower than 1250 K, and it is found to have a negative temperature dependence, which is on the contrary to the experimental result. Our results well capture the trend of the rate constant and temperature.

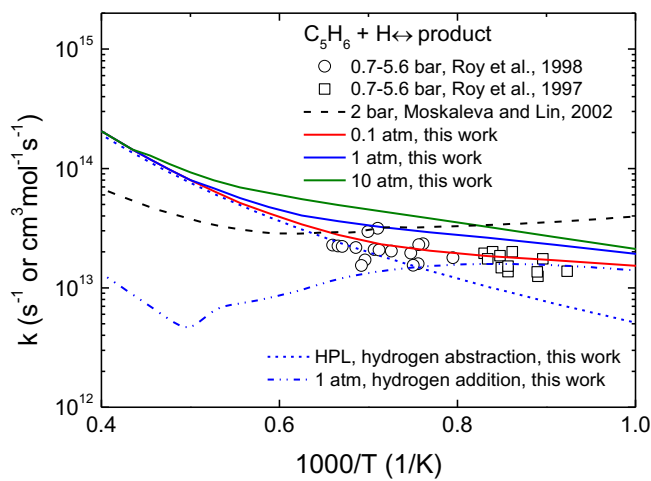

Figure. 8. Arrhenius plot of the total reaction constants of $C_5H_6$ + H from experiments at 0.7-5.6 bar [17,55], previous study at 2 bar [24] and the present study from 0.1 to 10 atm.

### 3.4 Dissociation reactions of $C_5H_7$ isomers

The dissociation reactions of $C_5H_7$ isomers are discussed in this section. Firstly, the high pressure limit rate constants of ring-opening reactions through C-C $\beta$-scission of cyclopentenyl radicals (W1, W2 and W3) to straight-chain $C_5H_7$ isomers (W4, W5, W6, W7) are illustrated in Figure 8. Rate constants of W1↔W4 and W2↔W6 from Zhong and Bozzelli [19] based on quantum chemistry and RRKM calculations, and that of W3↔W5 and W3↔W7 are estimated [60] based on the correlations proposed by Heyberger et al. [61] are also included in Figure 9 for comparison. The pressure dependent rate constants of the above dissociation reaction are detailed in the Supplementary Materials, which all present obvious fall-off effects at high temperatures. The dissociation rate constants of W1 to W4 is slightly lower than that of W2 to W6 as the barrier height of the two transition states are 44.58 and 43.6 kcal/mol, respectively. However, the HPL of the above two reactions are much lower than the rate constants provided by Zhong and Bozzelli [19] by orders of magnitude, especially for the dissociation of W1 to W4. It is about one order of magnitude smaller than that from Zhong and Bozzelli [19] at 2500 K, and it increases to about 6 orders of magnitude at 800 K. Such large difference comes from the underestimating the barrier height of T14 resulting from the difference in the prediction of barrier height of the isomerization from W1 to the linear RSR W3. That is, the barrier height by CCSD(T)/CBS//M06-2X/6-311+G(d, p) is 26.6 kcal/mol higher than that predicted using PM3 method, which has been discussed in Section 3.1. For the disscoaiton reactions from vinylic cyclopentenyl (W3) to W5 and W7, their HPL rate constants are similar, but at least two orders of magnitude larger than that of W1↔W4 and W2↔W6. The estimated rate constants [61] are roughly one order of magnitude smaller than the that from our current study.

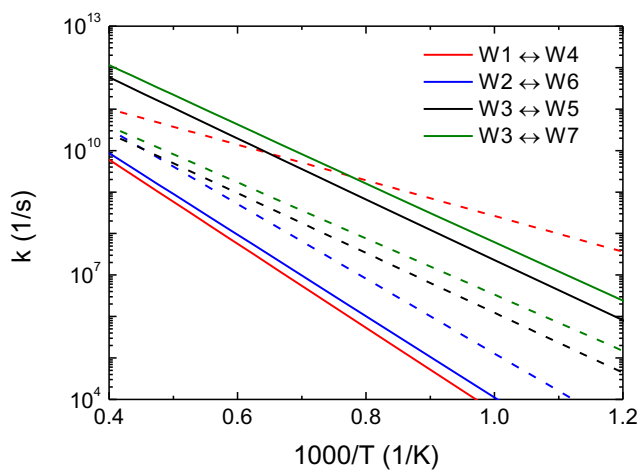

Figure 9. High pressure limit rate coefficients of dissociation reaction of cyclopentenyl to straight-chain C$_5$H$_7$ isomers, the soild lines are from present study and the the dashed lines are from literatures (W1↔W4 and W2↔W6[19], W3↔W5 and W3↔W7[60]). The color of the lines are related to specified reactions.

Afterwords, the dissociation reaction of straight-chain C$_5$H$_7$ isomers (W4, W5, W6, W7) and the branched C$_5$H$_7$ isomers (W9, W10) as discussed and the total dissociation rate constants of high pressure limit are illustrated in Fig. 10. For RSR W4, only the C-H fission is feasible, and that for branched RSR W9 is the C-C fission. However, For W5, W6 and W7, there are both C-C and C-H fissions, and for W10, there exits two kinds of C-C fission. For RSR W4 and W9, their dissociation channels are the slowest among their isomers. The predicted dissociation rate constants increase from W7, W5 to W6, in accordance with their dissociation thresholds displayed in Fig. 2. The difference is about one order of magnitude in the ranges of temperature and pressure of interest. For the dissociation of W10, it is at the same magnitude of W7, W5 and W6.

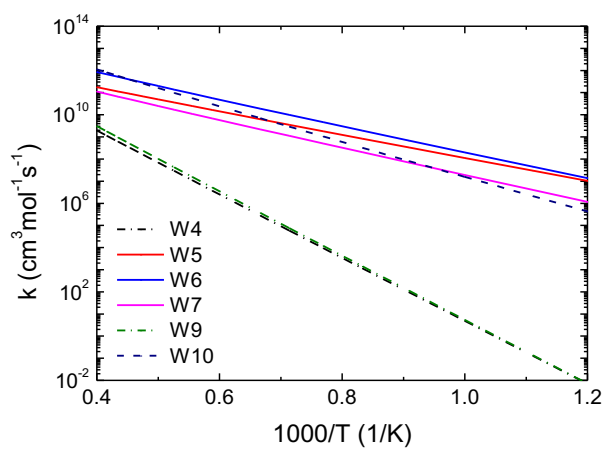

Figure 9. High pressure limit rate coefficients of the thermal decomposition rate constants of straight-chain C$_5$H$_7$ isomers (W4, W5, W6, W7) and branched C$_5$H$_7$ isomers (W8, W9)

The decomposition reaction of C$_5$H$_7$ isomers via C-C fission has relatively lower barrier height compared to the C-H fission as indicated by the PES in Fig. 2, and they are vital for fuel consumption. The rate coefficients of the major exit channels of W5 and W6 through C-C cleavage by forming biomolecular products of $C_2H_4$ and $C_3H_3$, $C_2H_2$ and A-$C_3H_5$ are presented in Fig. 10. Other dissociation reactions of straight-chain and branched $C_5H_7$ isomers are listed in the Supplementary Materials. Generally, the dissociation rate increases with the increasing pressure. Obvious fall-off effect was observed for the dissociation of the above two $C_5H_7$ isomers at high temperaures. Because of the well merging effect, the dissociation rate constants are missing at low temperature and low pressure regions. The high pressure dissociation rate constants of W5 dissociation estimated by [60] based on the correlations proposed by Heyberger et al. [61] is close to the result calculated based on current study, with a derivation less than two times. However, for the dissociation reaction of W6 to $C_2H_2$ and A-$C_3H_5$, the HPL rate coefficient provided by Zhong and Bozzelli [19] based on the semi-empirical method has a derivation of one order of magnitude smaller at 2500 K and becomes one order of magnitude larger at 800 K. This discrepancy originates from both energy differences and the hindered rotor treatment. As shown in Table 2, the barrier height at the level of PM3 is about 5.3 kcal/mol lower than that at the CCSD(T)/CBS level, while for the treatment of hindered rotors of the straight-chain W6, the torsions were assumed to have identical hindrance potential. However, as depicted in Fig. 3, they vary notably with molecular structures. In the current study, the hindrance potential for each rotor was carefully examined.

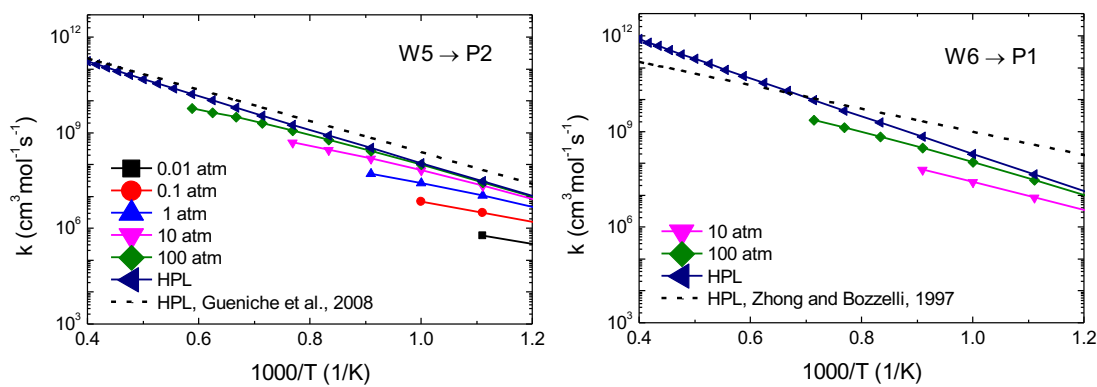

Figure 10. Pressure-dependent rate constants of W5 and W6 dissociation via C-C fission

## 4. Kinetic modeling

Reaction rates on $C_5H_6$ + H PES from the current study were incorporated into the ITV Mech [2,23] to further explore the effects of the missing reaction pathways as well as the pressure-dependent reaction kinetics on model prediction in several reported combustion experiments [15,16], such as premixed laminar flame speed of cyclopentadiene flames[15], the pyrolysis and oxidation of cyclopentadiene[16]. Details of the kinetic model and modeling have been provided in Section 2.3.

### 4.1 Premixed laminar flame speed

Firstly, the laminar flame speeds of cyclopentadiene/air mixture are simulated and compared against experimental result from Ji et al. [15], which were determined in a counterflow configuration at atmospheric pressure and for a wide range of equivalence ratios of $0.7 \leq \phi \leq 1.5$ at $T_0$ = 353 K. Figure 11 shows the results of laminar flame speeds of cyclopentadiene/air flames in experimental [15] and model prediction using ITV Mech as well as the updated ITV Mech. It is noteworthy that the updated ITV Mech improves the model prediction especially for fuel-rich conditions, while that for the fuel lean region, it is still within experimental uncertainty. The reaction pathway of CPD flames under fuel rich conditions were argued to result in substantial formation of propargyl and acetylene, exerting a strong influence on flame propagration [41].

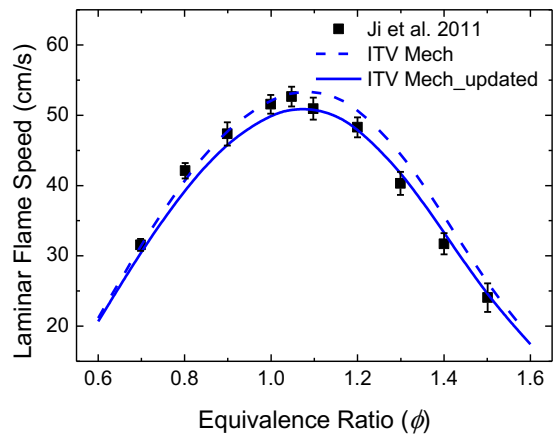

Figure 11. Premixed laminar flame speeds of cyclopentadiene/air flames at $T_0$ = 353 K and $P$ = 1 atm.

Square symbols: experimental data from Ji et al. [15]; Dash line: model prediction using ITV_Mech; Dashed line: model prediction using ITV_Mech_updated.

Afterwards, the logarithmic sensitivity coefficients of the laminar flame speed in lean ($\phi = 0.7$), stoichiometric ($\phi = 1$) and rich ($\phi = 1.4$) flame conditions at 1 atm and $T_0 = 353$ K were reported in Fig. 12 based on both the ITV Mech and the ITV Mech_updated. Notably, minor differences of the sensitive reaction sets for both models at the above three conditions. That is, the chain branching reaction of $H + O_2 = O + OH$ has the most sensitive positive effect on the laminar flame speed, while that of CPDyl and H recombination reaction has the most sensitive negative effect. The reactions involving the hydrogen addition and abstraction of CPD by hydrogen radicals both have negative effect on the laminar flame speed, which leads to the decrease of the laminar flame speed in the updated model shown in Fig. 11. The chain initiation reaction of $C_5H_6 = C_5H_5 + H$ and the chain propagation reaction of $H + O_2 + M = HO_2 + M$ retard the laminar flame speed and their impact are less important at fuel rich conditions. On the contrary, the chain termination reaction of $HCO + H = CO + H_2$ become more negative sensitive to the laminar flame speed with the increasing equivalent ratio. The chain initiation reaction of HCO and chain propagation of CO and OH both have positive effect on the laminar flame speed but opposite equivalence ration dependency.

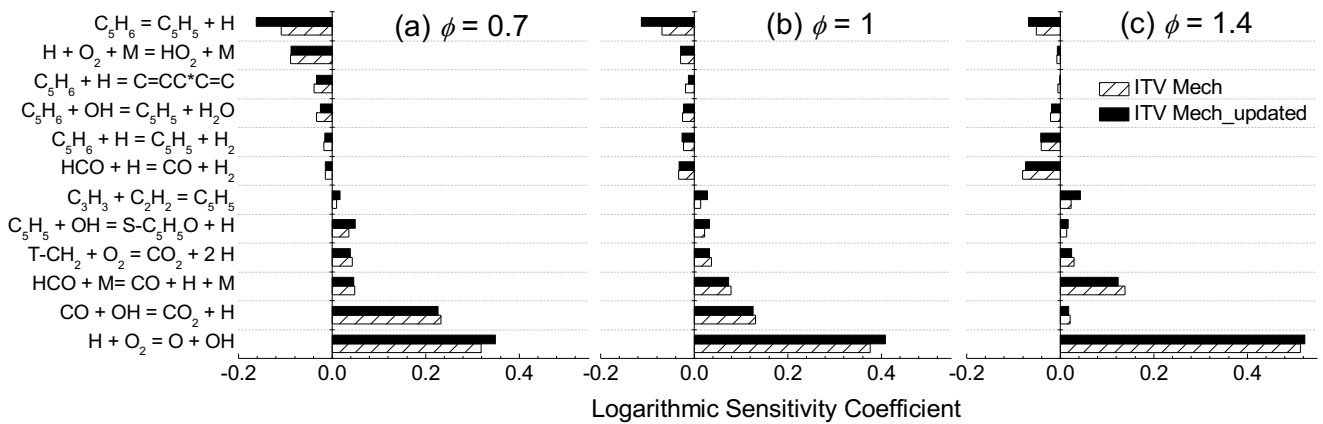

Figure 12. Logarithmic sensitivity coefficients of laminar flame speeds of cyclopentadiene/air flames

at $T_0$ = 353 K and $P$ = 1 atm, and fuel-air equivalence ratio of (a) $\phi$ = 0.7, (b) $\phi$ = 1, and (c) $\phi$ = 1.4.

## 4.2 Pyrolysis and oxidation of $C_5H_6$

To further explore the effects of the $C_5H_6$ + H kinetics on model prediction, both the pyrolysis and oxidation conditions of $C_5H_6$ were studied based on the original and updated ITV Mech and compared against experiments performed at varied concentration, equivalence ratio and initial temperature in the Princeton's adiabatic, atmospheric pressure flow reactor [16]. Here the predicted mole fraction profiles were shifted by 20 ms to match the fuel conversion in experiments, which was employed a usual practice to eliminate the experimental uncertainties in the zero-time specification [57]. Figure 12 and 13 compare the species mole fraction profiles predicted using both the original and updated ITV Mech and the experimental measurements in the pyrolysis condition at $\phi$ = 97.8, $T_0$ = 1202 K, and the oxidation condition at $\phi$ =1.03, $T_0$ =1198 K, respectively. Generally, the predicted mole fractions agree reasonably well with the experimental measurements especially in oxidation conditions. Significant improvement in the model prediction of the $C_2$ and $C_3$ species were observed by incorporating the updated $C_5H_6$ + H reactions for both conditions, whereas the predictions of benzene and naphthalene changes slightly. It is notably that the benzene is well captured in both the original and updated ITV Mech, while the naphthalene is under-predicted in pyrolysis conditions, which, however, has an opposite trend in the pyrolysis condition in the oxidation condition.

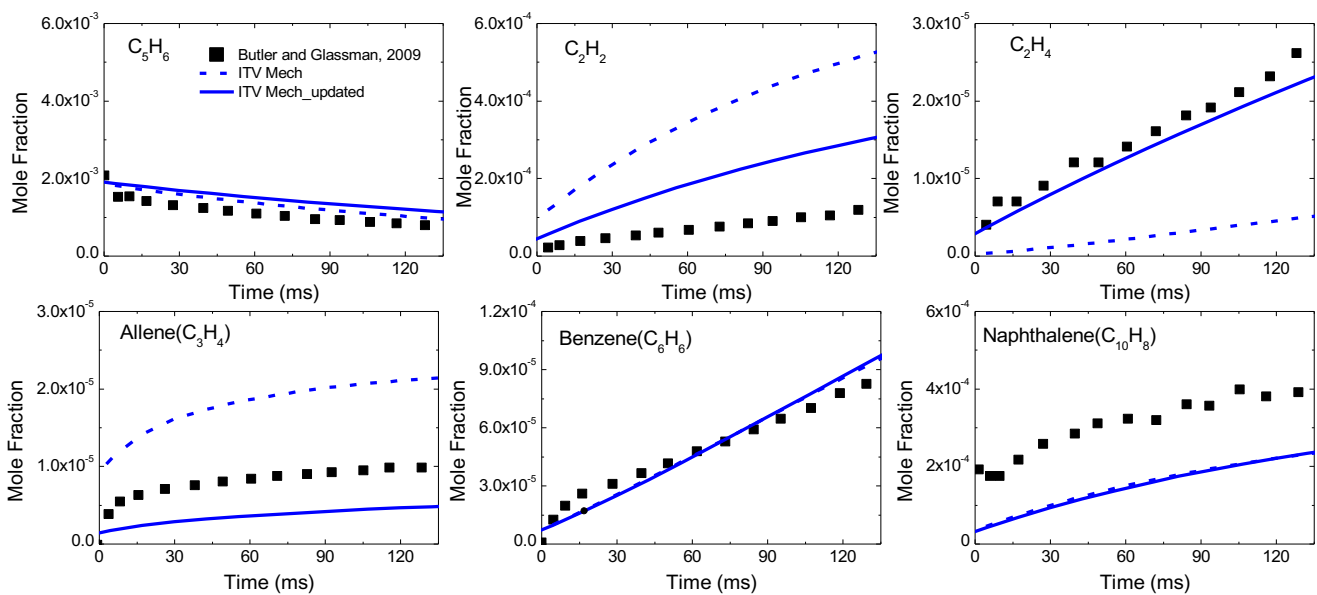

Figure 13 Experimental (symbols) and modeled (lines) mole fraction profiles of $C_5H_6$ pyrolysis in the Princeton plug flow reactor with $\phi = 97.8$, $T_0 = 1202$ K and $P = 1$ atm. The predicted mole fraction profiles were shifted by 20 ms. Dashed lines are original model results, while the solid lines represent the updated models incorporating the pressure-dependent rate constants computed in this work.

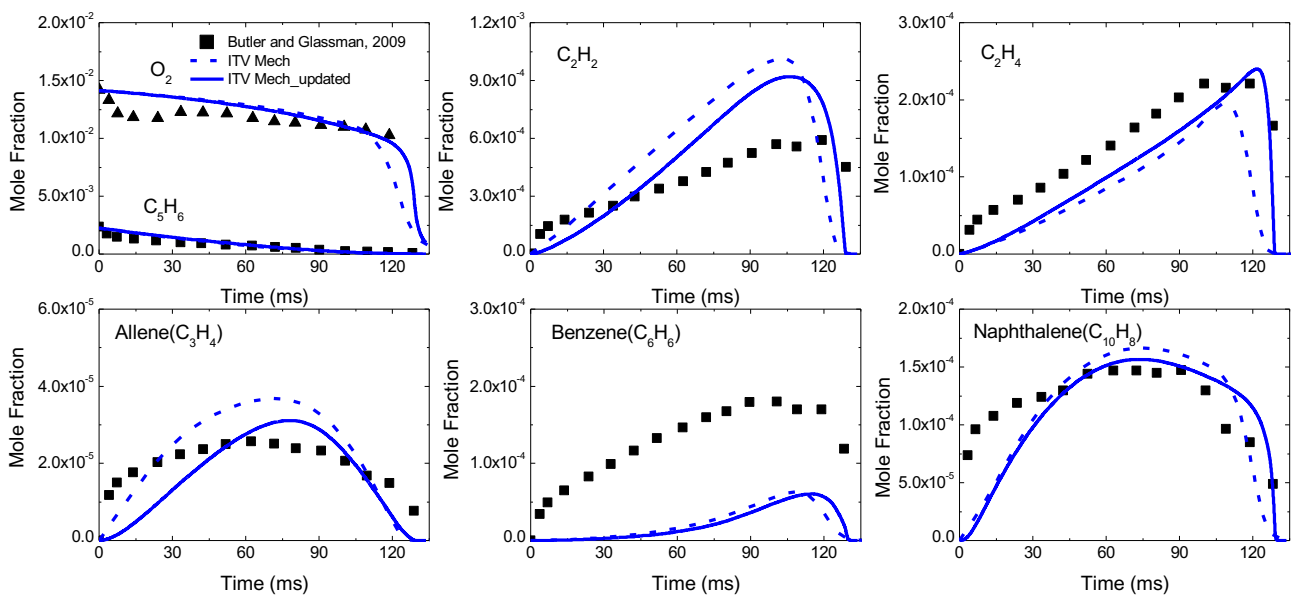

Figure 14 Experimental (symbols) and modeled (lines) mole fraction profiles of C5H6 oxidation in the Princeton plug flow reactor with $\phi = 1.03$, $T_0 = 1198$ K and $P = 1$ atm. The predicted mole fraction profiles were shifted by 20 ms. Dashed lines are original model results, while the solid lines represent the updated models incorporating the pressure-dependent rate constants computed in this work.

## 4. Conclusions

Cyclopentadiene (CPD) is an important intermediate in the combustion of fuel and formation of aromatics. Specifically, the reaction kinectis of H atom and CPD including abstraction and addition play significant roles in the polycyclic aromatic hydrocarbons (PAH) formation and fuel consumption. In the present study, the kinetics and thermodynamic properties for hydrogen abstraction and addition with CPD, and the related reactions including the isomerization and decomposition on the $C_5H_7$ potential energy surface were systematically investigated by theoretical calculations. High-level ab initio calculations were adopted to obtain the stationary points on the potential energy surfaces of CPD + H. Phenomenological rate coefficients for temperature- and pressure-dependent reactions in the full potential energy surface were calculated by solving the time-dependent multiple-well RRKM/master equation, and the hydrogen abstraction reactions were based on the conventional transition state theory. In terms of the hydrogen abstraction reactions, the hydrogen abstraction from the saturated carbon atom in CPD is found to be the dominant channel. For the hydrogen addition and the associated reactions on the $C_5H_7$ PES, the allylic and vinylic cyclopentenyl radicals and $C_2H_2 + C_3H_5$ were found to be the most important channels and reactivity-promoting products, respectively. Contributions of the addition and abstraction channels were assessed, and the results show that the hydrogen addition reactions are relatively favored over hydrogen abstraction reactions at low temperatures and high pressures. The previously neglected role of open-chain intermediates in the evaluation of the reaction kinetics has been suggested and the corresponding rate constants have been recommended for inclusion in the modeling of the H+ c-C5H6 reaction. Results indicate that the transformation from to straight-chain $C_5H_7$ is kinetically unfavorable due to the high strain energy of the 3-membered ring structure of the isomerization transition state. Moreover, the thermodynamic data and the calculated rate coefficients for both H atom abstraction and addition were incorporated into the ITV Mech to examine the impact of the computed pressure-dependent kinetics of $C_5H_6 + H$ reactions on model predictions. Significant improvement in the laminar flame speed of cyclopentadiene was observed, especially in the fuel-rich conditions. In addition, a noticeable improvement was found for the mole fractions of important intermediate species, e.g., acetylene, ethylene, allene, in both of the CPD pyrolysis and oxidation conditions.

## Acknowledgments

Support from the

## References

[1]  Marinov NM, Pitz WJ, Westbrook CK, Castaldi MJ, Senkan SM. Modeling of aromatic and polycyclic aromatic hydrocarbon formation in premixed methane and ethane flames. vol. 116–117. 1996. doi:10.1080/00102209608935550.

[2]  Baroncelli M, Felsmann D, Hansen N, Pitsch H. Investigating the effect of oxy-fuel combustion and light coal volatiles interaction: A mass spectrometric study. Combust Flame 2019;204:320–30. doi:10.1016/j.combustflame.2019.03.017.

[3]  Berta P, Aggarwal SK, Puri IK. An experimental and numerical investigation of n-heptane/air counterflow partially premixed flames and emission of NOx and PAH species. Combust Flame 2006;145:740–64. doi:10.1016/j.combustflame.2006.02.003.

[4]  Cypres R, Bettens B. Mecanismes de fragmentation pyrolytique du phenol et des cresols. Tetrahedron 1974;30:1253–60. doi:10.1016/S0040-4020(01)97298-9.

[5]     Manion JA, Louw R. Rates, products, and mechanisms in the gas-phase hydrogenolysis of phenol between 922 and 1175 K. J Phys Chem 1989;93:3563–74. doi:10.1021/j100346a040.

[6]     McEnally CS, Pfefferle LD. An experimental study in non-premixed flames of hydrocarbon growth processes that involve five-membered carbon rings. Combust Sci Technol 1998;131:323–44. doi:10.1080/00102209808935766.

[7]     Frenklach M. Reaction mechanism of soot formation in flames. Phys Chem Chem Phys 2002;4:2028–37. doi:10.1039/b110045a.

[8]     Wang Y, Chung SH. Soot formation in laminar counterflow flames. Prog Energy Combust Sci 2019;74:152–238. doi:10.1016/j.pecs.2019.05.003.

[9]     Miller JA, Klippenstein SJ. The Recombination of Propargyl Radicals and Other Reactions on a C6H6 Potential 2003:7783–99. doi:10.1021/jp030375h.

[10]    Melius, Carl F., Michael E. Colvin, Nick M. Marinov, William J. Pit and SMS. Reaction mechanisms in aromatic hydrocarbon formation involving the C5H5 cyclopentadienyl moiety. Symp Combust 1996;26:685–92.

[11]    Moskaleva L V., Mebel AM, Lin MC. The $CH_3$ + $C_5H_5$ reaction: A potential source of benzene at high temperatures. Symp Combust 1996;26:521–6. doi:10.1016/S0082-0784(96)80255-4.

[12]    Kislov V V., Mebel AM. The formation of naphthalene, azulene, and fulvalene from cyclic C 5 species in combustion: An ab initio/RRKM study of 9-H-fulvalenyl (C5H5-C5H4) radical rearrangements. J Phys Chem A 2007;111:9532–43. doi:10.1021/jp0732099.

[13]    Kislov V V., Mebel AM. An ab initio G3-type/statistical theory study of the formation of indene in combustion flames. II. The pathways originating from reactions of cyclic C 5 species-cyclopentadiene and cyclopentadienyl radicals. J Phys Chem A 2008;112:700–16.

doi:10.1021/jp077493f.

[14] Sharma S, Green WH. Computed rate coefficients and product yields for c-C5H 5 + CH3→ products. J Phys Chem A 2009;113:8871–82. doi:10.1021/jp900679t.

[15] Ji C, Zhao R, Li B, Egolfopoulos FN. Propagation and extinction of cyclopentadiene flames. Proc Combust Inst 2013;34:787–94. doi:10.1016/j.proci.2012.07.047.

[16] Butler RG, Glassman I. Cyclopentadiene combustion in a plug flow reactor near 1150 K. Proc Combust Inst 2009;32 I:395–402. doi:10.1016/j.proci.2008.05.010.

[17] Roy K, Frank P. High temperature pyrolysis and oxidation of cyclopentadiene. 21st Intl. Symp. Shock Waves, 1997, p. 403–7.

[18] Tsang W, Hampson RF. Chemical Kinetic Data Base for Combustion Chemistry. Part I. Methane and Related Compounds. J Phys Chem Ref Data 1986;15:1087. doi:10.1063/1.555759.

[19] Zhong X, Bozzelli JW. Thermochemical and kinetic analysis on the addition reactions of H, O, OH, and HO2 with 1,3 cyclopentadiene. Int J Chem Kinet 1997;29:893–913. doi:10.1002/(SICI)1097-4601(1997)29:12<893::AID-KIN2>3.0.CO;2-H.

[20] Zhong X, Bozzelli JW. Thermochemical and Kinetic Analysis of the H, OH, HO 2 , O, and O 2 Association Reactions with Cyclopentadienyl Radical. J Phys Chem A 1998;102:3537–55. doi:10.1021/jp9804446.

[21] Wang H, You X, Joshi A V, Davis SG, Laskin A, Egolfopoulos F, et al. USC mech version II. High-Temperature Combust React Model H 2007;2:96.

[22] Zhou CW, Li Y, O'Connor E, Somers KP, Thion S, Keesee C, et al. A comprehensive experimental and modeling study of isobutene oxidation. Combust Flame 2016;167:353–79. doi:10.1016/j.combustflame.2016.01.021.

[23] Narayanaswamy K, Pitsch H, Pepiot P. A chemical mechanism for low to high temperature oxidation of methylcyclohexane as a component of transportation fuel surrogates. Combust Flame 2015;162:1193–213. doi:10.1016/j.combustflame.2014.10.013.

[24] Moskaleva L V., Lin MC. Computational study of the kinetics and mechanisms for the reaction of H atoms with c-$C_5H_6$. Proc Combust Inst 2002;29:1319–27. doi:10.1016/S1540-7489(02)80162-6.

[25] Miller JA, Klippenstein SJ. Dissociation of propyl radicals and other reactions on a $C_3H_7$ potential. J Phys Chem A 2013;117:2718–27. doi:10.1021/jp312712p.

[26] Huang C, Yang B, Zhang F. Pressure-dependent kinetics on the $C_4H_7$ potential energy surface and its effect on combustion model predictions. Combust Flame J 2017;181:100–9. doi:10.1016/j.combustflame.2017.01.031.

[27] Narayanaswamy K, Blanquart G, Pitsch H. A consistent chemical mechanism for oxidation of substituted aromatic species. Combust Flame 2010;157:1879–98. doi:10.1016/j.combustflame.2010.07.009.

[28] Zhao Y, Truhlar DG. The M06 suite of density functionals for main group thermochemistry, thermochemical kinetics, noncovalent interactions, excited states, and transition elements: Two new functionals and systematic testing of four M06-class functionals and 12 other function. Theor Chem Acc 2008;120:215–41. doi:10.1007/s00214-007-0310-x.

[29] Alecu IM, Zheng J, Zhao Y, Truhlar DG. Computational thermochemistry: Scale factor databases and scale factors for vibrational frequencies obtained from electronic model chemistries. J Chem Theory Comput 2010;6:2872–87. doi:10.1021/ct100326h.

[30] Zheng, J., Alecu, I.M., Lynch, B.J., Zhao, Y. and Truhlar DG. Database of Frequency Scale Factors for Electronic Model Chemistries,Version 3 Beta 2. 2017.

[31] Rienstra-Kiracofe JC, Allen WD, Schaefer HF. C2H5+O2 reaction mechanism: High-level ab initio characterizations. J Phys Chem A 2000;104:9823–40. doi:10.1021/jp001041k.

[32] Gao LG, Zheng J, Fernández-Ramos A, Truhlar DG, Xu X. Kinetics of the Methanol Reaction with OH at Interstellar, Atmospheric, and Combustion Temperatures. J Am Chem Soc 2018;140:2906–18. doi:10.1021/jacs.7b12773.

[33] Truhlar DG. Basis-set extrapolation. Chem Phys Lett 1998;294:45–8. doi:10.1016/S0009-2614(98)00866-5.

[34] Frisch M, Trucks GW, Schlegel HB, Scuseria GE, Robb MA, Cheeseman JR, et al. Gaussian 09, revision D. 01 2009.

[35] Truhlar DG. A simple approximation for the vibrational partition function of a hindered internal rotation. J Comput Chem 1991;12:266–70. doi:10.1002/jcc.540120217.

[36] Pfaendtner J, Yu X, Broadbelt LJ. The 1-D hindered rotor approximation. Theor Chem Acc 2007;118:881–98. doi:10.1007/s00214-007-0376-5.

[37] Eckart C. The penetration of a potential barrier by electrons. Phys Rev 1930;35:1303–9. doi:10.1103/PhysRev.35.1303.

[38] Li X, You X, Law CK, Truhlar DG. Kinetics and branching fractions of the hydrogen abstraction reaction from methyl butenoates by H atoms. Phys Chem Chem Phys 2017;19:16563–75. doi:10.1039/c7cp01686g.

[39] Tardy DC, Rabinovitch BS. Collisional energy transfer. Thermal unimolecular systems in the low-pressure region. J Chem Phys 1966;45:3720–30. doi:10.1063/1.1727392.

[40] Troe J. Theory of thermal unimolecular reactions at low pressures. I. Solutions of the master equation. J Chem Phys 1977;66:4745–57. doi:10.1063/1.433837.

[41] Robinson RK, Lindstedt RP. On the chemical kinetics of cyclopentadiene oxidation. Combust

Flame 2011;158:666–86. doi:10.1016/j.combustflame.2010.12.001.

[42] Georgievskii Y, Miller JA, Burke MP, Klippenstein SJ. Reformulation and solution of the master equation for multiple-well chemical reactions. J Phys Chem A 2013;117:12146–54. doi:10.1021/jp4060704.

[43] Miller JA, Klippenstein SJ. Master equation methods in gas phase chemical kinetics. J Phys Chem A 2006;110:10528–44. doi:10.1021/jp062693x.

[44] Miller JA, Klippenstein SJ. From the multiple-well master equation to phenomenological rate coefficients: Reactions on a C3H4 potential energy surface. J Phys Chem A 2003;107:2680–92. doi:10.1021/jp0221082.

[45] Klippenstein SJ, Miller JA. From the time-dependent, multiple-well master equation to phenomenological rate coefficients. J Phys Chem A 2002;106:9267–77. doi:10.1021/jp021175t.

[46] Chase Jr MW. JANAF thermochemical tables third edition. J Phys Chem Ref Data 1985;14.

[47] Benson SW. Thermochemical kinetics: methods for the estimation of thermochemical data and rate parameters. New York, NY: Wiley; 1976.

[48] Goldsmith CF, Magoon GR, Green WH. Database of small molecule thermochemistry for combustion. J Phys Chem A 2012;116:9033–47. doi:10.1021/jp303819e.

[49] Al Rashidi MJ, Mármol JC, Banyon C, Sajid MB, Mehl M, Pitz WJ, et al. Cyclopentane combustion. Part II. Ignition delay measurements and mechanism validation. Combust Flame 2017;183:372–85. doi:10.1016/j.combustflame.2017.05.017.

[50] Pitsch H. FlameMaster: A C++ computer program for 0D combustion and 1D laminar flame calculations. Cited In 1998;81.

[51] Nunes PM, Agapito F, Cabral BJC, Borges Dos Santos RM, Martinho Simões JA. Enthalpy of

formation of the cyclopentadienyl radical: Photoacoustic calorimetry and ab initio studies. J Phys Chem A 2006;110:5130–4. doi:10.1021/jp060325n.

[52] Tian Z, Fattahi A, Lis L, Kass SR. Cycloalkane and cycloalkene C-H bond dissociation energies. J Am Chem Soc 2006;128:17087–92. doi:10.1021/ja065348u.

[53] Sharma S, Harper MR, Green WH. Modeling of 1,3-hexadiene, 2,4-hexadiene and 1,4-hexadiene-doped methane flames: Flame modeling, benzene and styrene formation. Combust Flame 2010;157:1331–45. doi:10.1016/j.combustflame.2010.02.012.

[54] Hou D, You X. Reaction kinetics of hydrogen abstraction from polycyclic aromatic hydrocarbons by H atoms. Phys Chem Chem Phys 2017;19:30772–80. doi:10.1039/C7CP04964A.

[55] Roy K, Horn C, Frank P, Slutsky VG, Just T. High-temperature investigations on the pyrolysis of cyclopentadiene. Symp Combust 1998;27:329–36. doi:10.1016/S0082-0784(98)80420-7.

[56] Emdee JL, Brezinsky K, Glassman I. A kinetic model for the oxidation of toluene near 1200 K. J Phys Chem 1992;96:2151–61. doi:10.1021/j100184a025.

[57] Djokic MR, Van Geem KM, Cavallotti C, Frassoldati A, Ranzi E, Marin GB. An experimental and kinetic modeling study of cyclopentadiene pyrolysis: First growth of polycyclic aromatic hydrocarbons. Combust Flame 2014;161:2739–51. doi:10.1016/j.combustflame.2014.04.013.

[58] Gao CW, Allen JW, Green WH, West RH. Reaction Mechanism Generator: Automatic construction of chemical kinetic mechanisms. Comput Phys Commun 2016;203:212–25. doi:10.1016/j.cpc.2016.02.013.

[59] Ristori A, Dagaut P, Bakali AEL, Pengloan G, Cathonnet M. Combustion Science and Technology Benzene Oxidation : Experimental Results in a JDR and Comprehensive Kinetic Modeling in JSR , Shock-Tube and Flame Benzene Oxidation : Experimental Results in a JDR

and Comprehensive Kinetic Modeling in JSR , Shock-Tube a. Combust Sci Technol 2001;167:223–56. doi:10.1080/00102200108952183.

[60] Gueniche HA, Glaude PA, Fournet R, Battin-Leclerc F. Rich methane premixed laminar flames doped by light unsaturated hydrocarbons: III. Cyclopentene. Combust Flame 2008;152:245–61. doi:10.1016/j.combustflame.2007.07.012.

[61] Heyberger B, Belmekki N, Erie VAL. Oxidation of Small Alkenes at High Temperature 2002. doi:10.1002/kin.10092.